\documentclass[preprint,aps,epsf]{revtex4}
\usepackage{graphicx,graphics} 

\def\s{\sigma}

\def\up{\uparrow}
\def\dd{\downarrow}
\def\las{\langle}
\def\ras{\rangle}

\begin{document}
\title{Spin-dependent resonant tunneling in ZnSe/ZnMnSe heterostructures}
\author{A. Saffarzadeh}
\altaffiliation[Corresponding author. ]{E-mail:
a-saffar@tehran.pnu.ac.ir} \affiliation{Department of Physics,
Tehran Payame Noor University, Fallahpour St., Nejatollahi St.,
Tehran, Iran}
\author{M. Bahar}
\affiliation{Department of Physics, Teacher Training University,
49 Mofateh Ave., Tehran, Iran}
\author{M. Banihasan}
\affiliation{Department of Physics, Islamic Azad University,
North Tehran Branch, Darband St., Tehran, Iran}

\begin{abstract}
Using the transfer matrix method and the effective-mass
approximation, the effect of resonant states on spin transport is
studied in ZnSe/ZnMnSe/ZnSe/ZnMnSe/ZnSe structures under the
influence of both electric and magnetic fields. The numerical
results show that the ZnMnSe layers, which act as spin filters,
polarize the electric currents. Variation of thickness of the
central ZnSe layer shifts the resonant levels and exhibits an
oscillatory behavior in spin current densities. It is also shown
that the spin polarization of the tunneling current in geometrical
asymmetry of the heterostructure where two ZnMnSe layers have
different Mn concentrations, depends strongly on the thickness
and the applied bias.
\end{abstract}

\maketitle

\newpage
\section{Introduction}
Tunneling and resonant tunneling processes involving electron
spin show prominent perspectives for new field spin-dependent
fast electronics. Besides important applications in new
spin-based multifunctional devices such as spin-field-effect
transistor and spin-light-emitting diode, the spin-dependent
resonant tunneling (SDRT) effect can also help us to more deeply
understand the role of spin degree of freedom of the tunneling
electron and the quantum size effects on spin transport processes
\cite{Wolf}. In recent years, some attempts have been made to
study the effects of SDRT in magnetic tunnel junctions in which
the electrons tunnel from one ferromagnetic metal electrode to
the other through a nonmagnetic metal layer and a thin insulator
layer \cite{Mood2,Sun,Lec1,Lec2,Yuasa}.

On the other hands, spin-injection into semiconductors by using
magnetic semiconductors has also been demonstrated, opening the
way to new "all-semiconductor" devices. II-VI diluted magnetic
semiconductors (DMSs) \cite{Schmidt1} are known to be good
candidates for effective spin injection into a nonmagnetic
semiconductor (NMS) because their spin polarization is nearly
100\% and their conductivity is comparable to that of typical
NMS. Moreover, II-VI DMSs can be n-type doped, thus avoiding the
very fast spin precession that limits the applicability of III-V
DMSs as spin injectors. A very promising II-VI DMS for spin
injection is (Zn,Mn)Se, which has been previously used for spin
injection experiments into GaAs \cite{Fiederling}, and ZnSe
\cite{Schmidt2}. More recently, Slobodskyy \emph{et al.}
\cite{Slobodskyy} using II-VI semiconductor layers, fabricated a
magnetic resonant tunneling diode. Their magnetic device is based
on a quantum well made of diluted magnetic semiconductor ZnMnSe
between two ZnBeSe barriers and surrounded by highly $n$-type
ZnSe layers.

Recently, Guo \emph{et al.} \cite{Guo1,Guo2} and Chang \emph{et
al.} \cite{Chang} investigated theoretically the spin-dependent
transport in ZnSe/Zn$_{1-x}$Mn$_x$Se DMS double barrier
structures. The results showed that the spin polarization of the
tunneling electrons can be tuned by changing the external
magnetic and electric fields. The effect of SDRT on spin
transport has also been studied in spin filter tunnel junctions
\cite{Filip,Saffar,Wilc,Shokri1,Shokri2} in which a ferromagnetic
semiconductor (FMS) such as EuS is used as a magnetic barrier.
The large spin polarization achievable using magnetic barriers
\cite{Mood3}, makes spin filtering a nearly ideal method for spin
injection into semiconductors, enabling novel spintronic devices
\cite{Fied}.

In Ref. \cite{Saffar}, using two FMS layers, we examined the SDRT
in a double barrier junction and in the coherent tunneling
regime. In the present work, we study theoretically the effects
of SDRT in ZnSe/ZnMnSe/ZnSe/ZnMnSe/ZnSe structures. The left and
right ZnSe layers are considered as emitter and collector
attached to external leads. We assume that the carrier wave
vector parallel to the interfaces and the carrier spin are
conserved in the tunneling process through the whole system. The
first assumption is well justified for interfaces between
materials whose lattice constants are nearly equal; and the
second one is also justified in our structure, because the sample
dimensions are much smaller than the spin coherence length. Using
the transfer matrix method and the nearly-free-electron
approximation we will study the effects of resonant tunneling on
spin-dependent current densities and the degree of spin
polarization.

The paper is organized as follows. In section II, the model is
described and then the spin current densities and spin
polarization for ZnSe/ZnMnSe heterostructures with two
paramagnetic layers, are formulated. In section III, the
numerical results for the symmetric and asymmetric structures are
discussed in terms of the thickness of the ZnSe layer, which has
been sandwiched between two paramagnetic layers. The results of
this work are summarized in section IV.

\section{Model and formalism}
We consider a spin unpolarized electron current injected into
ZnSe/Zn$_{1-x}$Mn$_x$Se/ZnSe/Zn$_{1-y}$Mn$_y$Se/ZnSe structures in
the presence of magnetic and electric fields along the growth
direction (taken as $z$ axis). The conduction electrons that
contribute to the electric currents, interact with the 3d$^5$
electrons of the Mn ions via the sp-d exchange interaction. Due
to the sp-d exchange interaction, the external magnetic field
gives rise to the spin splitting of the conduction band states in
the paramagnetic layers. Therefore, the injected electrons see a
spin-dependent potential. In the framework of the parabolic-band
effective mass approximation, the one-electron Hamiltonian of
such system can be written as
\begin{equation}\label{H}
H=\frac{1}{2m^*}({\bf P}+e{\bf
A})^2+V_s+V_0(z)+V_{\s_z}(z)-\frac{eV_az}{L} \ ,
\end{equation}
where the electron effective mass $m^*$ is assumed to be
identical in all the layers, and the vector potential is taken as
${\bf A}=(0,Bx,0)$. Here,
$V_s=\frac{1}{2}g_s\mu_B{\bf\s}\cdot{\bf B}$ describes the Zeeman
splitting of the conduction electrons, where ${\bf\s}$ is the
conventional Pauli spin operator. $V_0(z)$ is the conduction band
offset under zero magnetic field, which is the difference between
the conduction band edge of the ZnMnSe layers and that of the ZnSe
layer; $V_{\s_z}(z)$ is the sp-d exchange interaction between the
injected electron and the Mn ions and can be calculated within
the mean field approximation. Hence, the sum of the two terms can
be written as
\begin{eqnarray}\label{U}
V_0(z)+V_{\s_z}(z)&=&[V_{0L}-N_0\alpha_L\s_zx_{eff}\las S_{zL}\ras]
\Theta(z)\Theta(L_1-z)\nonumber\\
&&+[V_{0R}-N_0\alpha_R\s_zy_{eff}\las S_{zR}\ras]\Theta(z-L_1-L_2)
\Theta(L_1+L_2+L_3-z)\ ,
\end{eqnarray}
where, $V_{0L(R)}$, $N_0\alpha_{L(R)}$ and $\las
S_{zL(R)}\ras=SB_S[5\mu_BB/k_B(T+T_{0L(R)})]$ are respectively
the conduction band offset, the electronic sp-d exchange constant
and the thermal average of $z$th component of Mn$^{2+}$ spin in
the left (right) DMS layer. Here, $B_S(x)$ is the Brillouin
function and $S=5/2$ is the spin of the Mn ions. $\s_z=\pm1/2$ (or
$\up,\dd$) are the electron spin components along the magnetic
field, $x_{eff}=x(1-x)^{12}$ and $y_{eff}=y(1-y)^{12}$ are the
effective Mn concentrations due to the antiferromagnetic Mn-Mn
coupling, while $x$ and $y$ are real Mn concentrations.
$\Theta(z)$ is the step function, $L_1$ and $L_3$ are respectively
the widths of left and right DMS layers, and $L_2$ is the width
of the middle ZnSe layer. The last term in Eq. (\ref{H}) denotes
the effect of an applied bias $V_a$ along the $z$ axis on the
system, where $L=L_1+L_2+L_3$ is the total length of the
considered structure along the growth direction. It is important
to note that, our sample dimensions are much smaller than the
spin coherent length in the semiconductors. Therefore we have
neglected the effects of spin-flip processes in the Hamiltonian
of the system.

Here, we would like to point out that, due to the absence of any
kind of scattering center for the electrons, the motion along the
$z$-axis is decoupled from that of the $x$-$y$ plane which is
quantized in the Landau levels with energies
$E_n=(n+1/2)\hbar\omega_c$, where $n=0,1,2,\cdots$ and
$\omega_c=eB/m^*$. In such case, the motion of electrons along the
$z$ axis can be reduced to a one-dimensional Schr\"odinger
equation
\begin{equation}\label{HH}
-\frac{\hbar^2}{2m^*}\frac{d^2\psi_{\s_z}(z)}{dz^2}
+\left[V_s+V_0(z)+V_{\s_z}(z)-\frac{eV_az}{L}\right]\psi_{\s_z}(z)
=E_z\psi_{\s_z}(z) \ ,
\end{equation}
where $E_z$ is the longitudinal energy of electrons traversing
the heterostructure.

The general solution to the above Schr\"odinger equation is as
follows:
\begin{equation}\label{psi}
\psi_{j\s_z}(z)=\left\{\begin{array}{cc}
A_{1\s_z}e^{ik_{1\s_z}z}+B_{1\s_z}e^{-ik_{1\s_z}z}, & z<0 ,\\
A_{2\s_z}{\rm Ai}[\rho_{\s_z}(z)]+B_{2\s_z}{\rm Bi}[\rho_{\s_z}(z)] , & 0<z<L_1 ,\\
A_{3\s_z}{\rm Ai}[\rho_{\s_z}(z)]+B_{3\s_z}{\rm Bi}[\rho_{\s_z}(z)] , & L_1<z<L_1+L_2 ,\\
A_{4\s_z}{\rm Ai}[\rho_{\s_z}(z)]+B_{4\s_z}{\rm Bi}[\rho_{\s_z}(z)] , & L_1+L_2<z<L ,\\
A_{5\s_z}e^{ik_{5\s_z}z}+B_{5\s}e^{-ik_{5\s_z}z}, & z>L ,\\
\end{array}\right.
\end{equation}
where the coefficients $A_{j\s_z}$ and $B_{j\s_z}$ (with $j$=1-5)
are constants which can be determined by the boundary conditions,
Ai[$\rho_{\s_z}(z)$] and Bi[$\rho_{\s_z}(z)$] are Airy functions,
and
\begin{equation}
k_{1\s_z}=\sqrt{2m^*(E_z-V_s)}/\hbar \ ,
\end{equation}
\begin{equation}
k_{5\s_z}=\sqrt{2m^*(E_z-V_s+eV_a)}/\hbar \ ,
\end{equation}
\begin{equation}
\rho_{\s_z}(z)=-\frac{L}{eV_a\lambda}\left[V_s+V_0(z)+V_{\s_z}(z)
-E_z-\frac{eV_az}{L}\right]\ ,
\end{equation}
with
\begin{equation}
\lambda=\left[\frac{-\hbar^2L}{2m^*eV_a}\right]^{1/3} \ .
\end{equation}

The wave functions and their first derivatives in the five
regions are matched at the interfaces between the regions. The
matching results in a system of equations, which can be
represented in a matrix form \cite{Allen},
\begin{eqnarray}
\left(\begin{array}{cc}
A_{1\s_z}\\B_{1\s_z}
\end{array}\right)
=M_{total}\left(\begin{array}{cc} A_{5\s_z}\\B_{5\s_z}
\end{array}\right) \ ,
\end{eqnarray}
where $M_{total}$ is the transfer matrix that connects the
incidence and transmission amplitudes and has the following form
\begin{equation}\label{M}
M_{total}=M_1^{-1}(0)M_2(0)M_2^{-1}(L_1)M_3(L_1)M_3^{-1}(L_2)
M_4(L_2)M_4^{-1}(L_3)M_5(L_3) \ .
\end{equation}
Here,
\begin{equation}
M_j(z_i)=\left(\begin{array}{cc}
\psi_j^+(z)&\psi_j^-(z)\\
\frac{d\psi_j^+(z)}{dz}&\frac{d\psi_j^-(z)}{dz}
\end{array}\right)_{z=z_i}  \\,
\end{equation}
where, $\psi_j^+(z)$ and $\psi_j^-(z)$ are respectively the first
and second term of the wave functions in each layer, without
considering their coefficients. Since there is no reflection in
region 5, the coefficient $B_{5\s_z}$ in Eq. (\ref{psi}) is zero
and the transmission coefficient of the spin $\s_z$ electron,
which is defined as the ratio of the transmitted flux to the
incident flux, can be written as
\begin{equation}
T_{\s_z}(E_z,B,V_a)=\frac{k_{5\s_z}}{k_{1\s_z}}
\left|\frac{1}{M_{total}^{11}}\right|^2 \ ,
\end{equation}
where $M_{total}^{11}$ is the left-upper element of the matrix
$M_{total}$ defined in Eq. (\ref{M}). It should be noted that the
transmission coefficients depend on the incident energy $E_z$,
the magnetic field $B$, the applied bias $V_a$ and the spin
orientation. The spin-dependent current density is connected with
the transmission coefficients via
\begin{eqnarray}\label{J}
J_{\s_z}(B,V_a)&=&J_0B\sum_{n=0}^\infty\int_{0}^{+\infty}T_{\s_z}(E_z,B,V_a)\nonumber\\
&&\times\{f[E_z+(n+\frac{1}{2})\hbar\omega+V_s]
-f[E_z+(n+\frac{1}{2})\hbar\omega+V_s+eV_a]\}dE_z \ ,
\end{eqnarray}
where $J_0=e^2/4\pi^2\hbar^2$ and
$f(E)=1/\{1+\exp[(E-E_F)/k_BT]\}$ is the Fermi-Dirac distribution
function in which $E_F$ denotes the emitter Fermi energy.

The degree of spin polarization for electrons traversing the
heterostructure is defined by
\begin{equation}\label{p}
P=\frac{J_\dd(B,V_a)-J_\up(B,V_a)}{J_\dd(B,V_a)+J_\up(B,V_a)} \ ,
\end{equation}
where $J_\up$ $(J_\dd)$ is the spin-up (spin-down) current
density.
\section{Results and discussion}
From Eq. (\ref{J}), we can evaluate numerically the spin-dependent
current densities as a function of the thickness of the central
ZnSe layer for the symmetric and the asymmetric structures,
depending on the Mn concentration in both paramagnetic layers. In
all of the numerical calculations we use $m^*=0.16$ $m_e$ ($m_e$
is the mass of the free electron), $T=4.2$ K, $B=4$ T,
$L_1=L_3=75$ nm, $g_s=1.1$, and $E_F=5$ meV. Figs. 1(a) and 1(b)
show the thickness dependence of the current densities for
spin-up and spin-down electrons respectively, in a symmetric
($x=y=0.05$) structure. In both figures,
$N_0\alpha_L=N_0\alpha_R=0.26$ eV, $V_{0L}=V_{0R}=0$ meV,
$T_{0L}=T_{0R}=1.7$ K \cite{Dai}. By applying an external
magnetic field, the sp-d exchange interaction gives rise to a
giant spin splitting which is much larger than the Zeeman
splitting of the conduction electrons. In such case, the
degeneracy of the spin-up and spin-down electron states is
lifted, thus each paramagnetic layer in our structure behaves as
a potential barrier for spin-up electrons and a quantum well for
spin-down ones. Consequently, the spin-up electrons see a
double-barrier potential while the spin-down ones see a
double-well potential. The barrier potential lowers the current
density of spin-up electrons. For these electrons, the central
ZnSe layer behaves as a quantum well, thus with increasing the
thickness of the central layer, the position of the resonant
states formed in the well varies and shifts to the lower energy
side. This leads to the oscillations of the current density which
decrease exponentially at each applied bias. Our studies have
also shown that, with increasing the applied voltage, the
transmission coefficients for the energies which are near $E_F$,
increase. Since in tunneling process at low temperatures, the
electrons with energy near $E_F$ carry most of the current, one
can see that the current density increases with the applied bias.
On the other hand, as it is clear in Fig. 1(b), the current
densities and the amplitude of oscillations for spin-down
electrons are much higher than for spin-up ones. This is because
the structure behaves as a double-well potential, hence the
transmission coefficients are large even at low energy region and
the resonant peaks become less sharp. It is important to note
that in the spin-down case the resonance is originated from the
above-well virtual states.

In Figs. 2(a) and 2(b) we have shown the thickness dependence of
the current densities for spin-up and spin-down electrons
respectively, in an asymmetric ($x=0.04$, $y=0.05$) structure.
Here, for $x=0.04$, $N_0\alpha_L=0.27$ eV, $V_{0L}=-3$ meV,
$T_{0L}=1.4$ K, and for $y=0.05$, $N_0\alpha_R=0.26$ eV,
$V_{0R}=0$ meV, $T_{0R}=1.7$ K \cite{Dai}. At low applied bias,
the current density and the amplitude of oscillations for spin-up
electrons slightly increase when the thickness of $L_2$
increases, while with increasing the applied bias a reverse
behavior appears. An important point to note is that the values
of current density for the spin-up electrons in the asymmetric
structure is three orders of magnitude higher than the symmetric
one, while in the spin-down case, the values do not change
considerably relative to those obtained in the symmetric
structure. However, the behavior of current density for the two
spin subbands is completely different in both structures as shown
in Figs. 1 and 2. In the asymmetric structure and in the
spin-down case, with increasing the applied bias and the
thickness of the central ZnSe layer, the current density
increases, which is important for possible technological
applications. This is a direct consequence of the effect of
conduction band offset and the difference in the magnetic
impurity concentrations in the paramagnetic layers.

In order to further see the effects of the resonant states in our
system, we have shown in Figs. 3 and 4 the degree of spin
polarization as a function of $L_2$ under several applied
voltages in the symmetric and the asymmetric structures,
respectively. One can easily see in Fig. 3 that, the output
current of the symmetric structure exhibits high values of spin
polarization, which can reach 100\% when $L_2$ increases.
Therefore, the results display obvious behavior of spin filtering
effect in the symmetric structure, which is independent of the
applied voltage. The spin polarization in the asymmetric
structure has been shown in Fig. 4. The results show that (in
contrast with Fig. 3) in the asymmetric structure the spin
polarization strongly depends on the applied voltage. In such
structure at fixed $L_2$ the high (low) values of spin
polarization correspond to low (high) applied voltages. As the
thickness $L_2$ increases, the degree of spin polarization for
the low voltages decreases slowly, while the amplitude of
oscillations increases. However, the system for the high voltages
exhibits a reverse behavior. The results once again indicate that
the spin-dependent resonant states which are responsible for the
oscillations in spin currents, strongly depend on the external
electric field and the geometrical structure.

Therefore, the obtained results in the asymmetric structures
clearly illustrate that the current density and hence the degree
of spin polarization can be tuned by changing the applied voltage
and/or the thickness of the central ZnSe layer.

\section{Summary}
Based on the transfer matrix method and the effective-mass
approximation, we have investigated the effect of resonant states
on spin transport in
ZnSe/Zn$_{1-x}$Mn$_x$Se/ZnSe/Zn$_{1-y}$Mn$_y$Se/ZnSe
heterostructures. The numerical results show that the symmetric
structure enables the generation of spin-polarized injection
currents, since the heterostructure for each value of the applied
bias and the thickness of the central ZnSe layer, filters out the
contribution of spin-up electrons in the total current density.
On the other hand, our numerical calculations indicate that the
asymmetric structure can be a good spin filter, if we adjust the
applied voltage and the thickness. Such behaviors are originated
from the enhancement and suppression in the spin-dependent
resonant states. These interesting features are relevant for
devising tunable spin-dependent electronic devices such as spin
switches, tunable lasers \cite{Gruber,Buyanova}, magnetic
resonant tunneling diodes \cite{Slobodskyy}, and also useful to
study fundamental effects involved in the spintronic field.

\newpage
\begin{figure}
\centering \resizebox{0.86\textwidth}{0.4\textheight}
{\includegraphics{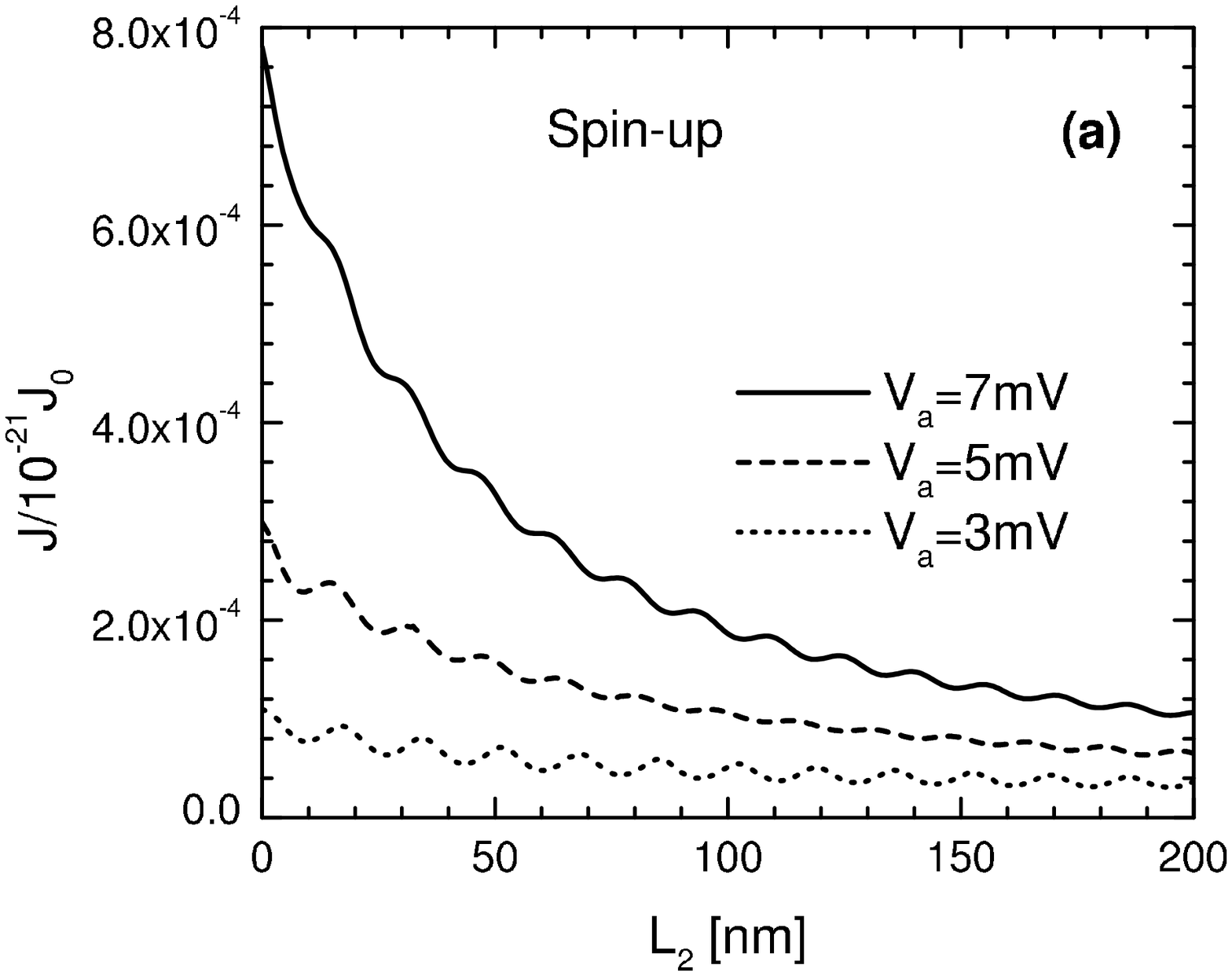}}
\centering\resizebox{0.78\textwidth}{0.4\textheight}
{\includegraphics{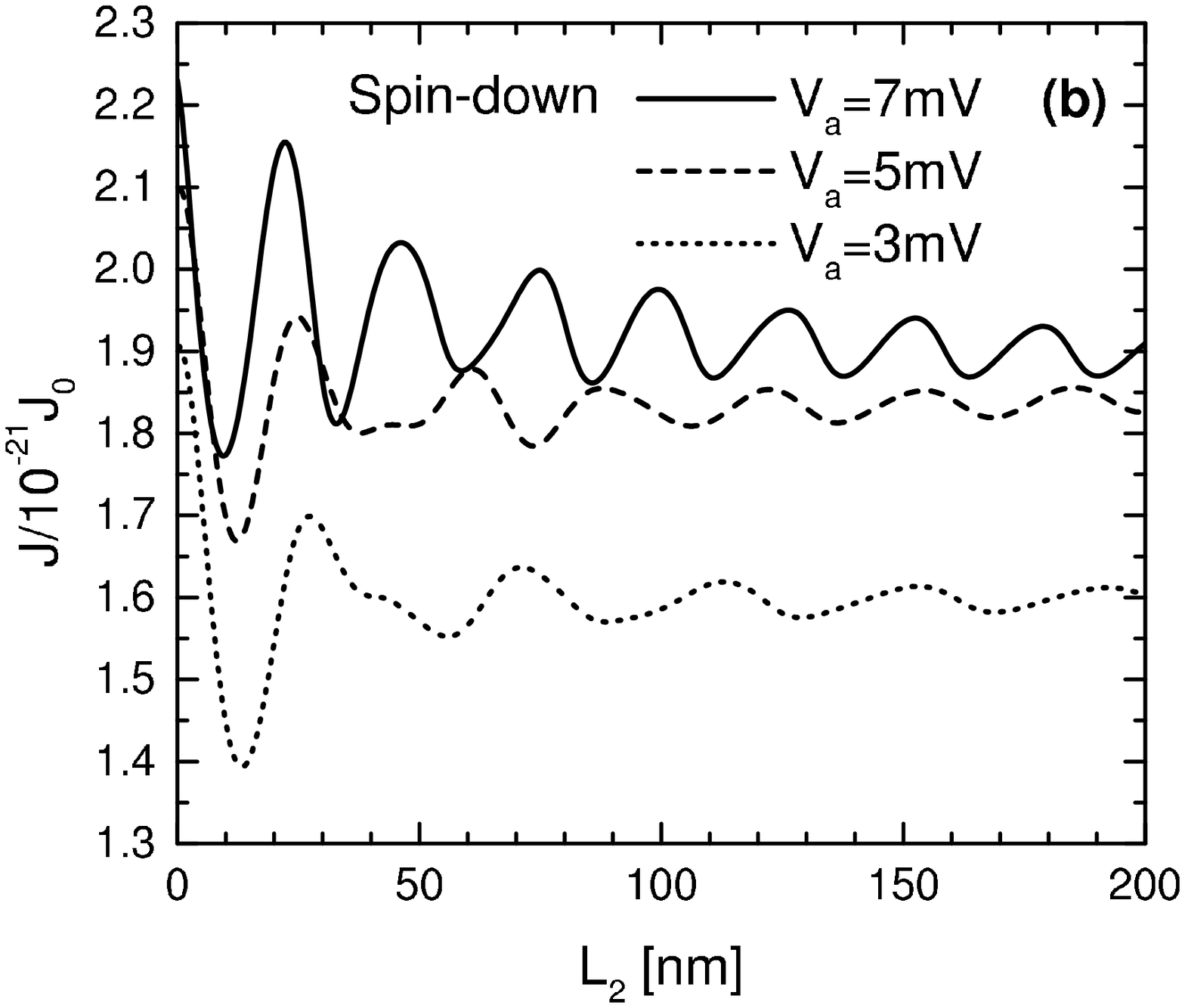}}\caption{(a) Spin-up and (b) spin-down
current densities as a function of the thickness $L_2$ for
ZnSe/Zn$_{1-x}$Mn$_x$Se/ZnSe/Zn$_{1-x}$Mn$_x$Se/ZnSe
heterostructures.}
\end{figure}

\newpage
\begin{figure}
\centering\resizebox{0.86\textwidth}{0.4\textheight}
{\includegraphics{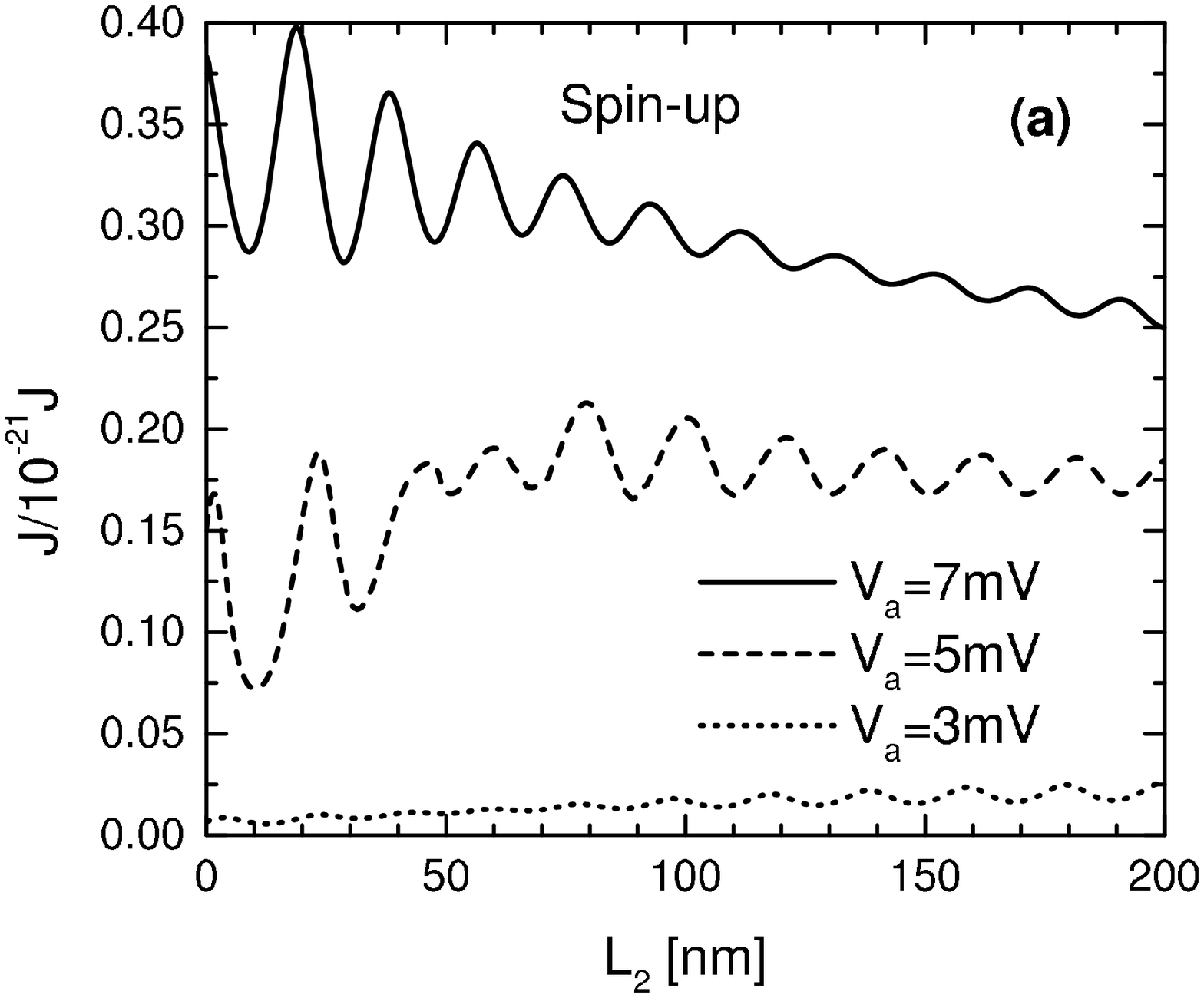}}
\centering\resizebox{0.84\textwidth}{0.4\textheight}
{\includegraphics{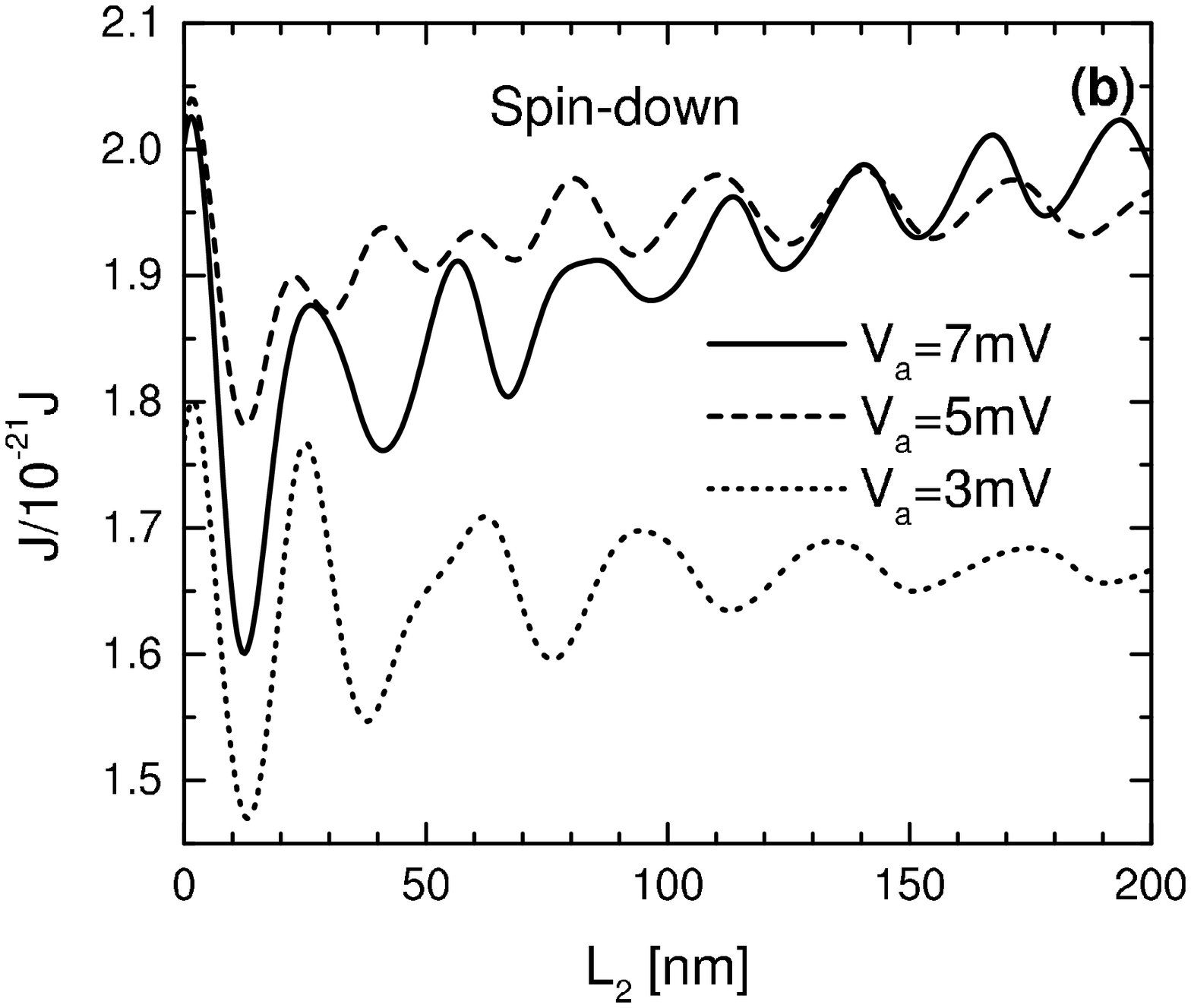}} \caption{(a) Spin-up and (b)
spin-down current densities as a function of the thickness $L_2$
for ZnSe/Zn$_{1-x}$Mn$_x$Se/ZnSe/Zn$_{1-y}$Mn$_y$Se/ZnSe
heterostructures.}
\end{figure}

\newpage
\begin{figure}
\centering \resizebox{0.86\textwidth}{0.4\textheight}
{\includegraphics{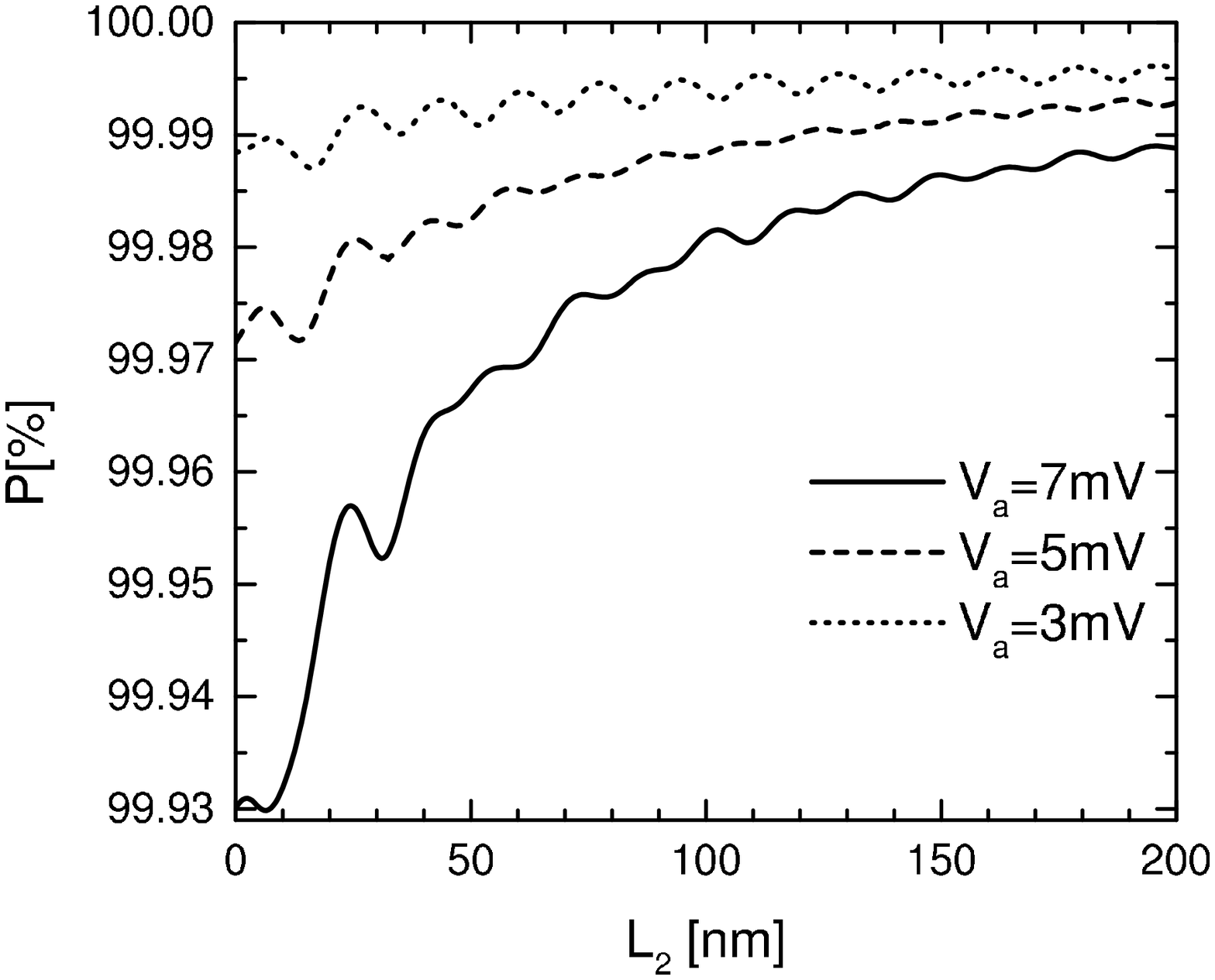}} \caption{Degree of spin polarization
as a function of the thickness $L_2$ for
ZnSe/Zn$_{1-x}$Mn$_x$Se/ZnSe/Zn$_{1-x}$Mn$_x$Se/ZnSe
heterostructures.}
\end{figure}

\newpage
\begin{figure}
\centering \resizebox{0.82\textwidth}{0.4\textheight}
{\includegraphics{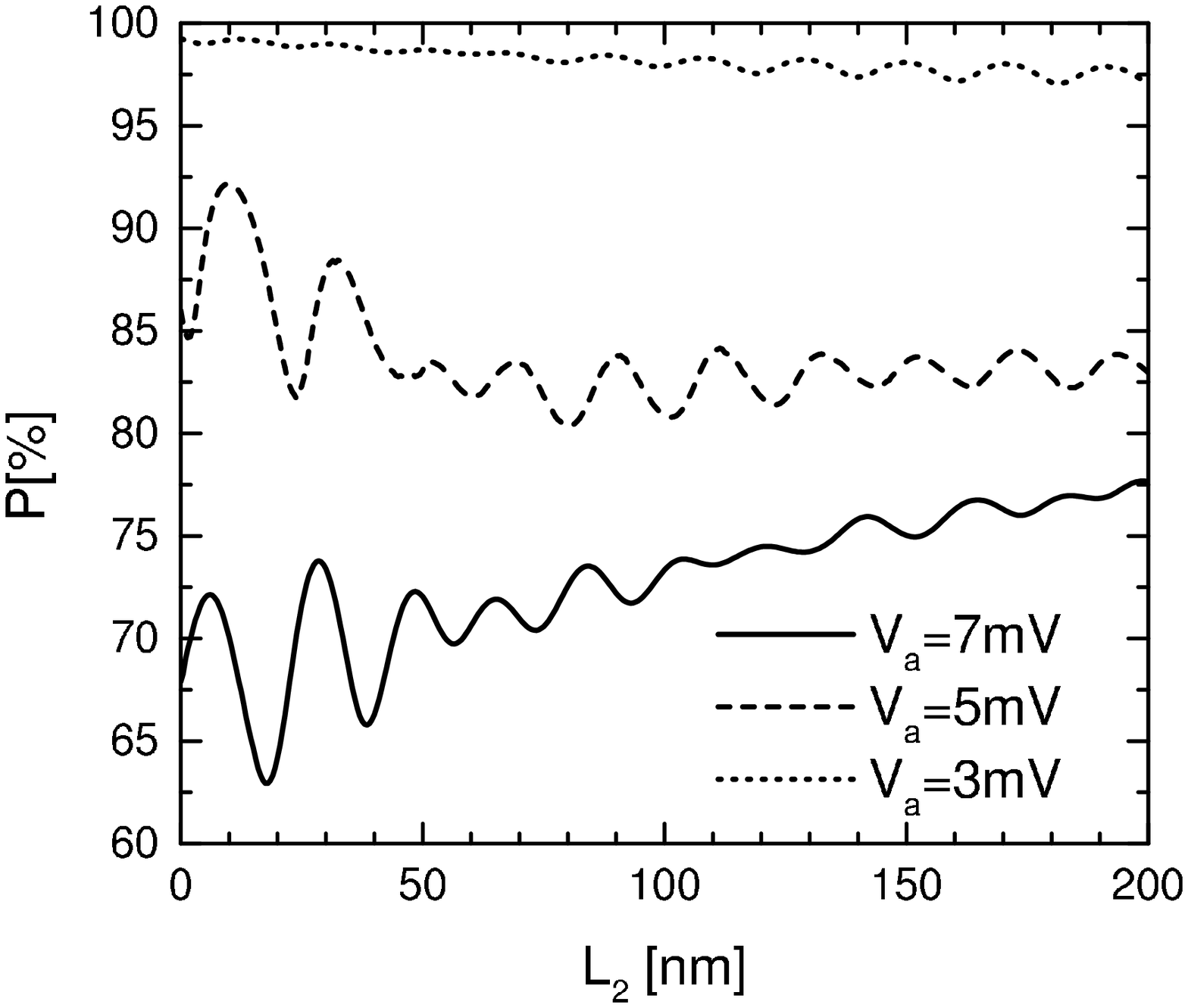}} \caption{Degree of spin polarization
as a function of the thickness $L_2$ for
ZnSe/Zn$_{1-x}$Mn$_x$Se/ZnSe/Zn$_{1-y}$Mn$_y$Se/ZnSe
heterostructures.}
\end{figure}
\end{document}